\documentclass[useAMS,usenatbib]{mn2e}
\usepackage{epsfig}
\usepackage{amsmath}

\newcommand{\be}{\begin{equation}}
\newcommand{\beq}{\begin{equation}}
\newcommand{\ba}{\begin{eqnarray}}
\newcommand{\ee}{\end{equation}}
\newcommand{\eeq}{\end{equation}}
\newcommand{\ea}{\end{eqnarray}}

\newcommand{\hs}{\hspace{1mm}}

\newcommand{\apj}{ApJ}
\newcommand{\aap}{A\&A}
\newcommand{\apjl}{ApJL}
\newcommand{\mnras}{MNRAS}
\newcommand{\aj}{AJ}

\newcommand{\nat}{{\it Nature}}
\newcommand{\araa}{ARA\&A}
\newcommand{\pasj}{PASJ}
\newcommand{\llya}{$L_{{\rm Ly}\alpha}$}

\def\lsim{~\rlap{$<$}{\lower 1.0ex\hbox{$\sim$}}}

\def\gsim{~\rlap{$>$}{\lower 1.0ex\hbox{$\sim$}}}

\title[The Impact of The IGM on High-Redshift Ly$\alpha$ Emission Lines]{The Impact of The IGM on High-Redshift Ly$\alpha$ Emission Lines}

\author[Mark Dijkstra, Adam Lidz \& Stuart Wyithe]{Mark Dijkstra$^{1}$\thanks{E-mail:dijkstra@physics.unimelb.edu.au}, Adam Lidz$^{2}$\thanks{E-mail:alidz@cfa.harvard.edu} and J. Stuart B. Wyithe$^{1}$\thanks{E-mail:swyithe@physics.unimelb.edu.au}\\
$^{1}$School of Physics, University of Melbourne, Parkville, Victoria, 3010, Australia\\
$^{2}$ Astronomy Department, Harvard University, 60 Garden Street, Cambridge, MA 02138, USA}

\def\LaTeX{L\kern-.36em\raise.3ex\hbox{a}\kern-.15em
    T\kern-.1667em\lower.7ex\hbox{E}\kern-.125emX}

\begin{document}

\date{\today}
\pagerange{\pageref{firstpage}--\pageref{lastpage}} \pubyear{2006}

\maketitle

\label{firstpage}
\begin{abstract}
We calculate the impact of the intergalactic medium (IGM) on the observed Ly$\alpha$ lines emitted by galaxies in an ionised IGM at $z\geq 4$. Our model accounts for gas clumping in the IGM and for the fact that high-redshift galaxies reside in overdense regions, which causes the velocity field of the IGM to depart from the Hubble flow. The observed shape of the Ly$\alpha$ line varies widely, with dependence on the intrinsic width and systemic velocity of the line, a galaxies star formation rate and the local extra-galactic UV-background. For large star formation rates and levels of the UV-background, absorption in the IGM does not result in a Ly$\alpha$ line that is asymmetric as is common among known high-redshift Ly$\alpha$ emitters. For models in which the lines do show the observed strong asymmetry, the IGM typically transmits only $10-30\%$ of the Ly$\alpha$ flux. The increase in the ionising background that accompanied the completion of reionisation barely increased the IGM transmission, which suggests that LAEs of comparable luminosity should not appear to be significantly dimmer prior to overlap. In this light, we briefly discuss the potential of Ly$\alpha$ emitters as a probe into the epoch of reionisation.
\end{abstract}

 
\section{Introduction}
\label{sec:intro}

The Ly$\alpha$ emission line is a powerful tracer of young galaxies \citep{PP67} and quasars \citep[e.g.][]{Fan06} in the high redshift Universe. Since the discovery of the first high-redshift Ly$\alpha$ emitters \citep[][hereafter, LAEs]{Hu96,Steidel96,Trager97}, existing Ly$\alpha$ surveys have detected several hundred galaxies out to redshift $z=7.0$ \citep[e.g.][]{Hu98,Rhoads00,MR02,Hu02,Ko03,Dawson04,Stanway04,Tran04,Taniguchi05,Westra06,Tapken06,Ka06,Shima06,Iye06,Stanway07}, and the number of these known LAEs is steadily growing. The total flux in the Ly$\alpha$ line emerging from galaxies provides a measure of its star formation rate, while its exact shape contain information on the sources powering the emission and on the gas distribution and kinematics through which the Ly$\alpha$ photons propagate. 

For example, the shape of the Ly$\alpha$ emission line associated with outflows produced by superwinds is typically peaked on a frequency redshifted by $v_{\rm w}$ relative to the true line center, where $v_{\rm w}$ is the speed of the outflowing gas. Furthermore, the spectrum is typically asymmetric and contains more photons redward of this peak \citep{Ahn04,Hansen06,Iro,V06}. On the other hand, the Ly$\alpha$ emission line from galaxies surrounded by neutral collapsing gas clouds is expected to be blueshifted relative to the true line center, by an amount that depends on the infall velocity and the total column density of neutral hydrogen in the infalling gas \citep{Zheng02,Iro,Dijkstra06a,V06}.

It is well known that the total flux in the Ly$\alpha$ line and its shape are affected by subsequent scattering in the IGM. To first order, the IGM at high redshift erases the blue half, and transmits the red half of the line, reducing the total Ly$\alpha$ flux by a factor of $2$. This is due to the fact that photons blueward of the Ly$\alpha$ transition will redshift into the Ly$\alpha$ resonance, where there is a large probability for the photon to be scattered out of the line of sight. Photons redward of the Ly$\alpha$ transition will redshift further and further away from resonance, which reduces the probability that these photons are scattered out of the line of sight. This simple picture would apply if the gas surrounding the galaxy is expanding with the Hubble flow. However, infall of IGM gas onto galaxies will occur out to distances well beyond the virial radius \citep{EL95,Barkana-Infall}. Furthermore, this infalling gas is expected to be significantly overdense, with values as large as $\Delta\sim 20$ times the mean density (just outside the virial radius \S~\ref{sec:IGM}).

Infall may have strong implications for the impact of the IGM on the Ly$\alpha$ emitted by galaxies. \citet{Santos04} has calculated the impact of an infalling neutral (i.e. pre-reionisation) IGM on a galaxies Ly$\alpha$ spectrum and found that the IGM typically erases a large fraction of the red side of the Ly$\alpha$ line. Therefore, much more than half of the Ly$\alpha$ is lost, with the IGM transmitting only $\mathcal{T}_{\alpha}\sim 10^{-3}-0.3$ of all Ly$\alpha$ photons. This low IGM transmission is partly due to the IGM damping wing, and partly due to resonant absorption by neutral hydrogen atoms in the infall region \citep[see Fig.~8 of][]{Santos04}. \citet{Barkana03} have found evidence for gas infall in the Ly$\alpha$ spectra of one bright quasar at $z=4.8$. Despite its large ionising flux, denser clumps in the infalling gas absorb observable amounts of Ly$\alpha$, reducing the total Ly$\alpha$ transmission well below $1$. In addition, \citet{Dijkstra06b} have shown that infall may also explain the observed spectrum of \citet{Steidel00}'s Ly$\alpha$ 'Blob' LAB1 \citep{Wilman05} and the peculiar line shape of a $z=5.8$ Ly$\alpha$ emitter observed by \citet{Bunker03}. These observations suggest that if infall is common around known LAEs, then their intrinsic Ly$\alpha$ luminosities, and consequently, their star-formation rates, may have been (significantly) underestimated in many cases, since the Ly$\alpha$ transmission is generally not corrected for infall.

In this paper we address the question: How does the IGM affect the shape and transmission of the Ly$\alpha$ emission line? Our approach is similar to the work of \citet{Santos04} and our results are consistent whenever there is overlap. However, we expand upon this work in several important ways: ({\it 1.}) we incorporate gas clumping in the IGM; ({\it 2.}) we investigate the impact of clustering of galaxies on the IGM transmission; ({\it 3}) we calculate the skewness parameter, $S_W$, which was recently introduced by \citet{Ka06}. However, the biggest difference between this paper and previous work is our focus on the reionised IGM: although it is generally understood that a neutral IGM can strongly suppress Ly$\alpha$ emission from galaxies \citep{GP,Haiman02,Santos04}, the impact of an ionised IGM has been virtually ignored. As discussed in \S~\ref{sec:z7} our analysis is also relevant in scenarios in which large disjoint HII bubbles exist prior to the end of reionisation \citep{F04a,WL04a}. Note that we focus on the Ly$\alpha$ line emitted by galaxies. For a brief discussion on the impact of the IGM on Ly$\alpha$ lines emitted by quasars, the reader is referred to \citet{agn}. The outline of this paper is as follows: We review the mechanism that creates Ly$\alpha$ emission in star forming galaxies, and discuss its expected properties in \S~\ref{sec:lya}. We describe our model for the IGM in \S~\ref{sec:IGM}. In \S~\ref{sec:result} we present calculations of the modified shape of the Ly$\alpha$ line. We discuss various implications of our results in \S~\ref{sec:discuss}, before presenting conclusions in \S~\ref{sec:conclusions}.  The parameters for the background cosmology used throughout this paper are $\Omega_m=0.24$, $\Omega_{\Lambda}=0.76$, $\Omega_b=0.044$, $h=0.73$ and $\sigma_8=0.74$ \citep{Spergel06}.

\section{Ly$\alpha$ Emission from Star Forming Galaxies }
\label{sec:lya}
\subsection{The Ly$\alpha$ Luminosity vs. Star Formation Rate}
\label{sec:sfr}
In this section we review the relation between the star formation
rate in galaxies and Ly$\alpha$ luminosity, and discuss its uncertainties. The Ly$\alpha$ flux from star forming galaxies originates
in the dense nebulae from which the stars form. Ionising photons produced by O stars, which are absorbed in the nebulae, are converted into Ly$\alpha$ in about $2$ out of $3$ cases \citep[][for case-B recombination, p 367]{Osterbrock}. Under the assumption that all Ly$\alpha$ photons escape from the galaxy into the IGM (sometimes referred to as the 'dust-uncorrected' Ly$\alpha$ luminosity), the total Ly$\alpha$ luminosity of a galaxy can then be estimated from 
\begin{equation}
L_{{\rm Ly}\alpha}=0.68 h\nu_{\alpha}[1-f_{\rm esc}] \dot{Q}_H,
\label{eq:llya}
\end{equation} where $h\nu_{\alpha}=10.2$ eV is the energy of a Ly$\alpha$ photon, $f_{\rm esc}$ is the escape fraction of ionising photons, and $\dot{Q}_H$ is total luminosity of ionising photons. As $\dot{Q}_H$ depends on the number of O-stars, its value depends on the assumed IMF and metallicity of the gas from which the stars form. Under the assumption of constant star formation rate (as opposed to a star burst), \citet{S03} has calculated $\dot{Q}_H$ for several different IMFs, and for a range of metallicities $Z$ (expressed in solar units $Z_{\odot}$). For a Salpeter IMF with lower and upper mass limits of $M_l=M_{\odot}$ and $M_u=100M_{\odot}$, respectively 
\begin{equation}
\log_{10} \dot{Q}_H=53.8+\log_{10}\dot{M}_*-0.0029(9+\log_{10}Z)^{2.5},
\label{eq:qdot}
\end{equation} in which $\dot{M}_*$ is the star formation rate in $M_{\odot}$/yr. \citet{S03} also calculates the time-evolution of $\dot{Q}_H$ for an instantaneous star burst. In this case, $\dot{Q}_H$ can be very high, reaching values of $\log_{10}\dot{Q}_H({\rm photons\hs s}^{-1} M_{\odot}^{-1})\sim 47-48$, during the first few Myr after the star burst. For the reminder of this paper, we will assume the star formation does not take place in such a burst and we will use Eq~(\ref{eq:llya}) and Eq~(\ref{eq:qdot}) to calculate \llya\hs as a function of $\dot{M}_*$. Following \citet{Santos04}, our models assume $f_{\rm esc}=0.1$ \citep{I06} and $Z=0.05Z_{\odot}$, unless stated otherwise.\footnote{For comparison, given solar metallicities ($Z=Z_{\odot}$) the total output of ionising photons is related to the star formation rate through $\log_{10} \dot{Q}_H=53.0+\log_{10}\dot{M}_*$ \citep{K98}. This relation is often used in the literature and is a lower by factor of $2$ than the relation used in this paper.  The uncertainties in the relation between $L_{{\rm Ly}\alpha}$ and $\dot{M}_*$ introduced by the gas metallicity and by the IMF are ignored for the reminder of this paper. Here, our focus is on the uncertainties introduced by the IGM.} This low metallicity reflects that galaxies at $z \gsim 4-5$ live in Universe that is $\lsim 1$ Gyr old, and in which there has been less time for their gas to become enriched by metals and dust. 

\subsection{The Shape+Width of the Intrinsic Ly$\alpha$ Line}
\label{sec:lyashape}

For simplicity, the shape of the intrinsic Ly$\alpha$ line is usually assumed to be Gaussian. For gas that is optically thin to Ly$\alpha$ photons, a reasonable choice for the standard deviation of this Gaussian emission line is $\sigma_{\rm \alpha}\sim 1.5 v_{\rm circ}$ \citep{Santos04}. However, to take this value for all galaxies assumes that each star forming, Ly$\alpha$ emitting galaxy is a disk seen edge-on, which is only true for a small subset of LAEs. For disk galaxies seen face-on, the Ly$\alpha$ line is narrower. To account for this inclination effect, the fiducial intrinsic Ly$\alpha$ emission line should be lowered to $\sigma_{\rm \alpha}\sim 0.7\times 1.5 v_{\rm circ}\approx v_{\rm circ}$ (where $\langle |\cos i|\rangle \sim 0.7$, and $i$ is inclination angle of the galaxy). 

As already mentioned, the previous discussion applies when the gas inside the galaxy is optically thin to Ly$\alpha$ photons. Since the line-center Ly$\alpha$ optical depth exceeds unity for hydrogen columns $N_H \gsim 10^{13}$ cm$^{-2}$, the optically thin approximation is most likely incorrect. Indeed, as mentioned in \S~\ref{sec:sfr} the bulk of Ly$\alpha$ photons are believed to be emitted inside dense nebulae in which star formation is occurring. For ionising photons to be converted into Ly$\alpha$, these nebulae must be optically thick to ionising radiation. This requires hydrogen to have a column of $N_H \gsim 10^{17}$ cm$^{-2}$, making the line center Ly$\alpha$ optical depth $\tau_0 \gsim 10^4$. Resonant scattering of Ly$\alpha$ photons through optically thick neutral hydrogen gas clouds may boost the width of the Ly$\alpha$ emission line. However, resonant scattering is not expected to boost the Ly$\alpha$ line width significantly.

For example, if we have Ly$\alpha$ photons emitted in the center of a static uniform slab/sphere with line-center optical depth $\tau_0$, then resonant scattering causes the emerging spectrum to comprise two peaks, each separated from the line center by $\sim 180(N_H/10^{20}\hs {\rm cm}^{-2})^{1/3}(T_{\rm gas}/10^4 \hs {\rm K})^{1/6}$ km s$^{-1}$, where $T_{\rm gas}$ is the gas temperature \citep{Harrington73,Neufeld90,Dijkstra06a}. The FWHM of the resonant profile is $\sim 2.5$ times as large as this separation. 

\begin{figure}
\vbox{\centerline{\epsfig{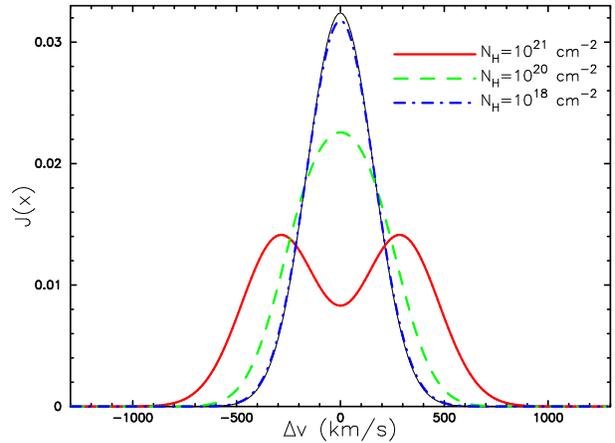}}}
\caption[]{Broadening of the Ly$\alpha$ emission line through resonant scattering. The Ly$\alpha$ flux density $J(x)$ (arbitrary units) is shown as a function of velocity off-set from the true line center. A Gaussian Ly$\alpha$ emission line with $\sigma_v=160$ km s$^{-1}$, was convolved with the 'resonant profile', which is the line profile that would emerge from a static, uniform slab with a hydrogen column density $N_{\rm H}$ (see text). When $N_{H}\gsim 10^{20}$ cm$^{-2}$, resonant scattering of the Ly$\alpha$ emission line becomes more important.}
\label{fig:emissionline}
\end{figure}

To quantify the impact of resonant scattering on the width of the line, we convolve a Gaussian emission line with the resonant profile derived by \citet{Harrington73} and \citet{Neufeld90}. For this example, the Gaussian has a standard deviation of $v_{\rm circ}=160$ km s$^{-1}$. The associated physical picture is that Ly$\alpha$ photons emerge from the nebula in which they were created with a double peaked emission spectrum, after which they can freely escape from the galaxy. For simplicity, we assume that the total column of hydrogen, $N_H$, is the same for each nebula. In Figure~\ref{fig:emissionline} we show Ly$\alpha$ emission lines for various values of $N_H$. The figure shows that broadening of the line is not important for $N_{H}\lsim 10^{20}$ cm$^{-2}$. The main reason is that in this example, the FWHM of the Gaussian is comparable to the FWHM of the resonant profile for $N_H=10^{20}$ cm$^{-2}$, which is $\sim 450$ km s$^{-1}$. For larger columns the Ly$\alpha$ line broadens and displays the double-peak of the resonant profile. 

The column densities of atomic hydrogen observed in local galaxies typically lie in the range $N_H=10^{20}-10^{21}$ cm$^{-2}$ \citep[e.g.][]{Bosma81,Kunth98}, but may reach larger values. This suggests that resonant scattering can boost the observed Ly$\alpha$ widths significantly. However, the above calculation represents an upper limit on the amount of resonant broadening for a given $N_H$, because the calculation assumed the hydrogen gas to be uniform and static. In reality, velocity gradients in the gas will facilitate the escape of Ly$\alpha$ from the nebula, resulting in a reduced broadening through resonant scattering. Furthermore, the observation that ionising radiation does emerge from star forming regions suggests that paths through the galaxy must exist along which $N_H \lsim 10^{17}$ cm$^{-2}$. If Ly$\alpha$ photons escape through these lower column-density paths, then resonant broadening of the Ly$\alpha$ line is negligible. This also applies when the ISM consists of two phases in which case resonant scattering is even less important (see \S~\ref{sec:ew}). The possibility that the majority of Ly$\alpha$ photons escape through these 'low column density paths', has been invoked as an explanation that the Ly$\alpha$ photons manage to escape from galaxies at all \citep[e.g][]{Neufeld91,Kunth98}.\\

We conclude that resonant scattering of Ly$\alpha$ photons is not expected to broaden the width of the Ly$\alpha$ emission line of galaxies with $v_{\rm circ} \gsim 150-200$ km s$^{-1}$ by a factor $\gsim 2$. For significantly less massive galaxies, i.e. $v_{\rm circ} \lsim 50$ km s$^{-1}$, resonant scattering could in principle broaden the line to a few hundred km s$^{-1}$ (since this requires columns of only $N_H\sim 10^{20}$ cm$^{-2}$) though it may not if the Ly$\alpha$ escapes through 'low column density paths'.

\subsection{The Equivalent Width of the Ly$\alpha$ Line}
\label{sec:ew}
If we approximate the spectrum of a source blueward of the Ly$\alpha$ frequency using a powerlaw of the form $f_{\nu}=k \nu^{\beta}$, then the maximum intrinsic Ly$\alpha$ EW is given by \citep[][for $\beta < -1$]{Charlot93}

\begin{equation}
{\rm EW_{\rm max}}=\frac{0.68 h\nu_{\alpha}\int_{\nu_H}^{\infty}\frac{f(\nu)}{h\nu}d\nu}{f_{\lambda}(\lambda=\lambda_{\alpha})}=-0.68\frac{\lambda_{\alpha}}{\beta}\Big{(}\frac{\nu_H}{\nu_\alpha}\Big{)}^{\beta}, 
\end{equation} where $h=6.62 \times 10^{-27}$ erg s, is Planck's constant and $\nu_H=3.28\times 10^{15}$ Hz, is the hydrogen ionisation threshold. The Ly$\alpha$ flux was calculated assuming that two-thirds of all ionising photons are converted into Ly$\alpha$, i.e. we assumed the escape fraction of ionising photons to be $f_{\rm esc}=0$. EW$_{\rm max}$ is then obtained after dividing this number by the continuum flux density (in erg s$^{-1}$ $\AA^{-1}$) at the Ly$\alpha$ frequency. Here, we used the relation $\lambda f_{\lambda}=\nu f_{\nu}$ and have denoted the Lyman limit frequency with $\nu_H$. For stars, the index $\beta$ is typically $-4$ to $-5$ and we get that ${\rm EW_{\rm max}}=64$ \AA\hs(for $\beta=-4$). \citet{Charlot93} showed, using stellar synthesis models in combination with realistic stellar spectra, that the maximum EW that is reached in the first $\sim 10^7$ yrs after the onset of star formation and can be as high as ${\rm EW_{\rm max}}\sim 300$ \AA, although the precise value depends on the assumed IMF \citep[e.g.][]{S03}.

 Resonant scattering of Ly$\alpha$ photons increases the total distance traveled before the photons emerge from the galaxy. Therefore, Ly$\alpha$ is more subject to dust attenuation than the continuum photons, which can result in a reduction of the Ly$\alpha$ EW. However, the opposite may be true in a multi-phase ISM \citep{Neufeld91,Hansen06}, in which case the dust is locked up in cold clumps embedded in a hot ionised medium. The Ly$\alpha$ photons can escape from the galaxy scattering off the dust clumps, while mainly traveling through the ionised interclump medium. Depending on the details of clump size, clump dust content, etc, the fraction of Ly$\alpha$ transmitted through the ISM could exceed that of continuum photons. Therefore, the intrinsic Ly$\alpha$ EW produced by a normal stellar population could, in some cases, be larger than $300$ \AA. 
\section{Modeling the Intergalactic Medium}
\label{sec:IGM}
 Our approach to modeling the IGM is based on \citet{Barkana-Infall} who calculated the average density and velocity profiles of intergalactic gas around virialised halos of various masses.  It is worth emphasising that infall of gas onto galaxies is inherent in the Press-Schechter model for the collapse of cold dark matter halos. This model is described below.
\subsection{The IGM Density and Velocity Field}
\label{sec:rho}
Figures 2 and 4 of \citet{Barkana-Infall} show the average radial density and velocity profile of the intergalactic gas beyond the virial radius $r_{\rm vir}$ of objects of mass $M_{\rm tot}$ at various redshifts. Here, we focus on the density profile associated with $z=5$ and $M_{\rm tot}=10^{10}M_{\odot}-10^{12}M_{\odot}$. Just outside the virial radius, the average density, $\rho(r)$ is $\rho(r_{\rm vir})=20\bar{\rho}$, where $\bar{\rho}$ is the mean baryon density of the universe. The density drops off approximately following a power-law $\rho(r)\propto r^{-1}$, and reaches a value of $\rho(r=10r_{\rm vir})=2\bar{\rho}$. At larger radii, the density asymptotically approaches $\bar{\rho}$ (see Fig~2 of Barkana, 2004). The velocity profile (Fig~4 of Barkana, 2004), shows the IGM to be collapsing at $\sim v_{\rm circ}$, just outside the virial radius. The IGM is stationary with respect to the galaxy at $r\sim 5r_{\rm vir}$. In this model, the IGM never becomes fully comoving, which is not surprising since the average density within a sphere of radius $r$ is always larger than $\bar{\rho}$. The model should therefore break down beyond a certain radius. In this paper, we assume that this model applies up to $10r_{\rm vir}$. Beyond this radius, we assume the IGM is simply comoving and at the mean density in the universe. To summarise, our fiducial IGM density and velocity fields are given by
\begin{equation}
\label{eq:rho}
\rho(r)=
\left\{ \begin{array}{ll}
         \ 20\bar{\rho}(r/r_{\rm vir})^{-1}& r < 10r_{\rm vir};\\

         \ \bar{\rho} & r \geq 10r_{\rm vir}\end{array} \right.
\end{equation}

\begin{equation}
v_{\rm IGM}(r)=
\left\{ \begin{array}{ll}
         \ -v_{\rm circ}+ \Big{(}\frac{r-r_{\rm vir}}{9r_{\rm vir}}\Big{)} \times \\
\ \Big{(}10r_{\rm vir}H(z) +v_{\rm circ}\Big{)}& r_{\rm vir} <r < 10r_{\rm vir};\\
         \ H(z)r & r \geq 10r_{\rm vir},\end{array} \right.\nonumber
\label{eq:vel} 
\end{equation} where $H(z)$ is the Hubble constant at redshift $z$.	
\subsection{Photoionisation of the IGM}
\label{sec:photoion}

If the galaxy spectrum is of the form $J(\nu) \propto \nu^{\beta}$ blueward of the hydrogen ionisation threshold, $\nu_H$, then the total photoionisation rate at distance $r$ from the galaxy is

\begin{equation}
\Gamma_{\rm ion}=B_{\rm c}(r)\Gamma_{\rm BG}+\frac{\beta}{\beta-3}\frac{\sigma_0\dot{Q}_H f_{\rm esc}}{4\pi r^2}, 
\label{eq:gamma}
\end{equation} where we approximate the hydrogen photoionisation cross section as $\sigma_H(\nu)=\sigma_0(\nu/\nu_H)^{-3}$, in which $\sigma_0=6.3\times 10^{-18}$ cm$^{-2}$. The photoionisation rate due the extra-galactic background, $\Gamma_{\rm BG}$, has been determined observationally to be $\sim 10^{-13}$ s$^{-1}$ between $z=5$ and $z=6$ \citep{Fan06}. The factor $B_{\rm c}(r)$ is the (radius dependent) boost in the ionising background due to the clustering of surrounding galaxies. Clustering is ignored in our fiducial model, i.e. $B_{\rm c}(r)=1$, but may be as large as $10^2$ (see \S~\ref{sec:clustering}). For a given $\Gamma_{\rm ion}$, the neutral fraction of hydrogen is calculated 
from
\begin{equation}
 x_H=1+\frac{\Gamma_{\rm ion}}{2n_{H}\alpha_{\rm rec}}-\Big{[}\frac{\Gamma_{\rm ion}}{n_{H}\alpha_{\rm rec}}+\Big{(}\frac{\Gamma_{\rm ion}}{2n_{H}\alpha_{\rm rec}}\Big{)}^2\Big{]}^{1/2}
\label{eq:xh}
\end{equation} which was derived assuming photoionisation equilibrium and ignoring helium. Here, $n_H$ is the number density of hydrogen nuclei and the case-A recombination coefficient $\alpha_{\rm rec}=4.2 \times 10^{-13}(T_{\rm gas}/10^4\hs{\rm K})^{0.7}$ cm$^3$ s$^{-1}$ \citep[e.g.][]{Hui97}. We have verified that the total optical depth to ionising radiation is negligible in our calculations. We note that our optically-thin approximation is invalid if a dense, self-shielded and optically thick Lyman-limit system (LLs) lies along a given sightline. If anything, the presence of LLs would lower the actual photoionisation rate, which would increase the neutral fraction of hydrogen, which would in turn lower the overall Ly$\alpha$ transmission.
\subsection{The Impact of Clustering of Sources on the Ionising Background}
\label{sec:clustering}
We have calculated the enhancement of the ionising background due to clustering of nearby galaxies using a semi-analytic model based on that described in \citet{WL05}. Details of the model can be found in Appendix~A, and we summarise only the its main features here. We define a mean free path of ionising photons by $\lambda$, which we assume to be independent of radius. At a distance $r$ from the central galaxy, this flux may be written as
\begin{equation}
\label{eq:boost}
F_{\rm bg}(r) = \int d^3x \frac{\frac{dN}{dt}\left(\left|\vec{x}\right|\right)}{4\pi\left|\vec{r}-\vec{x}\right|^2}\exp{\left[-\frac{\left|\vec{r}-\vec{x}\right|}{\lambda}\right]},
\end{equation} where $dN/dt(r)$ gives the ionising photon production rate per unit volume from sources surrounding the central galaxy at radius $r$. The model computes $dN/dt(r)$ using a radius dependent mass function for galaxies in the infall-region. More details are given in Appendix~A (Eq~\ref{rate}). Note that several model assumptions enter the expression of $dN/dt(r)$, which introduces some uncertainty into the exact enhancement of the background flux. However, the main purpose of this calculation, is to show that the boost in the ionising background is large (this statement is independent of the detailed model assumptions).

\begin{figure}
\vbox{\centerline{\epsfig{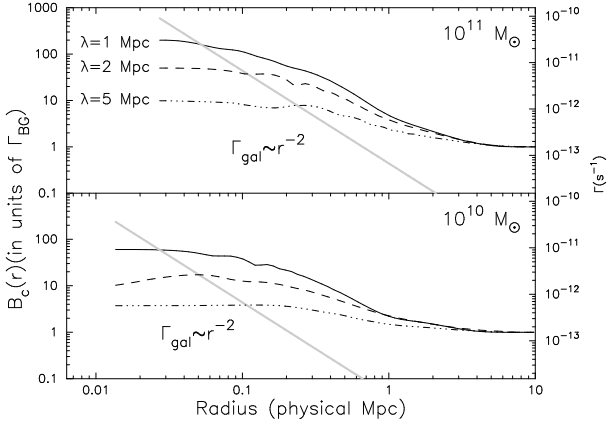}}}
\caption[]{The boost in the ionising background $B_{\rm c}(r)$ due to clustered sources surrounding the central galaxy for the cases $M_{\rm tot}=10^{11}M_{\odot}$ ({\it top panel}) and $M_{\rm tot}=10^{10}M_{\odot}$ ({\it bottom panel}). The boost is shown for 3 values of the assumed mean free path of the ionising photons (text). The vertical axis on the right shows the associated physical photoionisation rate (in s$^{-1}$). For comparison, the photoionisation rate due to the central galaxy is shown as the {\it thick grey line} for $\dot{M}_*= 10^2 M_{\odot}$/yr ({\it top panel}) and $\dot{M}_*=10 M_{\odot}$/yr ({\it bottom panel} also in units of the measured $\Gamma_{\rm BG}$). For both cases, $B_{\rm c}(r) \rightarrow 10^2$ when $\lambda \lsim 2$ Mpc. Clustering of sources may boost the ionising background significantly. As a result, the total photoionisation rate is not dominated by the central galaxy beyond a few virial radii.}
\label{fig:prox}
\end{figure}

The resulting ionising background is shown in Figure~\ref{fig:prox} for the cases in which the central galaxy sits in dark matter halos of mass $M_{\rm tot}=10^{11}M_{\odot}$ ({\it top panel}) and $M_{\rm tot}=10^{10}M_{\odot}$ ({\it bottom panel}). The flux $F_{\rm bg}(r)$ is written in units of the mean background flux [i.e. we have plotted $B_{\rm c}(r)$, Eq~(\ref{eq:gamma})]. To show the dependence of the boost on the assumed $\lambda$, $B_{\rm c}(r)$ is shown for $\lambda=1,2$ and $5$ physical Mpc, which covers the range of $\lambda$ derived by \citet{Fan06}. Note that $B_{\rm c}(r) \rightarrow 10^2$ for $\lambda \leq 2$ Mpc in both cases. For comparison, the ionising flux from the central galaxy is shown as the {\it thick grey line} for $\dot{M}_*=10^2 M_{\odot}$/yr ({\it top panel}) and $\dot{M}_*= 10 M_{\odot}$/yr ({\it bottom panel}, more precisely, $\Gamma_{\rm gal}/\Gamma_{\rm BG}$ is shown, see Eq~\ref{eq:gamma}). In all cases, the total ionising flux is not dominated by that of the central galaxy beyond a few virial radii. The impact of this enhancement on the Ly$\alpha$ line is discussed in \S~\ref{sec:param}.

\subsection{The IGM Transmission}
\label{sec:igmtrans}

The total opacity seen by a photon initially at frequency $\nu$ is
\begin{equation}
\tau(\nu)=\sigma_0 \int_{r_{\rm vir}}^{\infty}ds  \hs n_{\rm H}(s) \hs x_{\rm H}(s)\hs \phi\Big{(}x(\nu)-\frac{v_{\rm IGM}(s)}{v_{\rm th}}\Big{)},
\label{eq:tauH}
\end{equation} where we denoted frequency in units of the dimensionless variable $x\equiv (\nu-\nu_{\alpha})/\Delta \nu_D$, in which $\nu_D=\nu_{\alpha}(v_{\rm th}/c)$; $v_{\rm th}=(2kT_{\rm gas}/m_p)^{1/2}$ is the thermal speed of atoms in the gas at temperature $T_{\rm gas}$. We assume the gas temperature to be $T_{\rm gas}=10^4$ K, which is appropriate for the photoionised IGM. Note that the temperature of the IGM may actually have been as high as $T_{\rm gas}=5\times 10^4$ K shortly after reionisation was completed \citep[e.g.][]{Hui03}. However, we found that our results are practically insensitive to the precise value of $T_{\rm gas}$. The Ly$\alpha$ absorption cross section is written as $\sigma(x)=\sigma_0 \phi(x)$, where $\sigma_0=5.67 \times 10^{-14}(T_{\rm gas}/10^4\hs{\rm K})^{-1/2}$ cm$^{2}$ and $\phi(x)$ is given by
\begin{equation}
\phi(x)=
\left\{ \begin{array}{ll}
         \ \sim  e^{-x^2}& \mbox{core};\\
         \ \sim \frac{a}{\sqrt{\pi}x^2}& \mbox{wing},\end{array}. 
\right. 
\label{eq:phi}
\end{equation} 
The gas density profile $n_H(r)$ given by Eq~(\ref{eq:rho}), is azimuthally averaged. In reality, for a given line of sight, the density fluctuates around this average. 
These density fluctuations can cause significant deviations from the opacity
calculated from Eq~(\ref{eq:tauH}). If the density fluctuations are characterised by a distribution $P(\Delta)d\Delta$, which gives the probability that the gas density at radius $r$ is in the range $\rho(r)[\Delta \pm d\Delta/2]$ (where $\rho(r)$ is given by Eq~\ref{eq:rho}), then the average IGM transmission at frequency $x$
\begin{equation}
\langle e^{-\tau} \rangle (x)=\int_0^{\Delta_{\rm max}}d\Delta\hs P(\Delta)\hs e^{-\Delta^2 \tau(x)}.
\label{eq:trans}
\end{equation}  This equation is equivalent to Eq~(9) in Barkana \& Loeb (2004). Here, we take $\Delta_{\rm max}=100$, but note that the results presented depend very weakly on its value. The function $P(\Delta)d\Delta$ is not well known.  Following Barkana \& Loeb (2004) we take the distribution $P(\Delta)d\Delta$ derived by \citet{M00}, who obtained this distribution at various redshifts from a simulation. This means that the density fluctuations in the overdense regions are assumed to be the same as those at mean density.

Since $\phi(x)$ is peaked strongly around $x=0$, Eq~({\ref{eq:tauH}) states that the largest contribution to $\tau(x)$ comes from gas that resonates with the photons at this frequency. By combining Eq~(\ref{eq:gamma}), Eq~(\ref{eq:xh}) and Eq~(\ref{eq:tauH}) we can calculate the impact of the IGM on the Ly$\alpha$ emission line. 
\section{Results}
\label{sec:result}
\begin{figure}
\vbox{\centerline{\epsfig{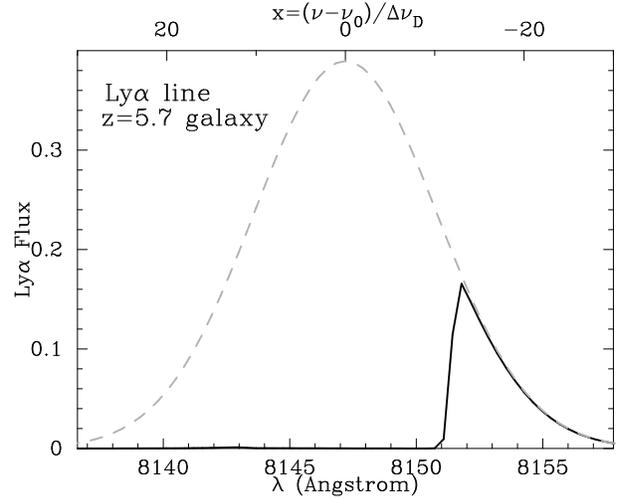}}}
\caption[]{The impact of a post-reionisation i.e. fully ionised IGM on the observed shape of Ly$\alpha$ emission line emitted by a $z=5.7$ star forming galaxy($\dot{M}_*=10M_{\odot}$/yr) in a dark matter halo with a mass of $M_{\rm tot}=10^{11}M_{\odot}$ (this mass is relevant for the extent of the infall region). The intrinsic shape ({\it dashed line}) of the Ly$\alpha$ line is Gaussian. The {\it solid} represents the observed shape after IGM processing. Here, only $\sim 12\%$ of the Ly$\alpha$ is transmitted to the observer.}
\label{fig:fig1}
\end{figure}
Our fiducial model consists of a star forming galaxy embedded in a dark matter halo with a total mass of $M_{\rm tot}=10^{11}M_{\odot}$. At $z=5.7$, the halos' associated virial radius and circular velocity are $23$ kpc and $133$ km s$^{-1}$, respectively. The value of $M_{\rm tot}$ is chosen to facilitate comparison with Santos (2004). However, this choice of $M_{\rm tot}$ may be justified by the observed bias and clustering of high-redshift LAEs which suggest $M_{\rm tot}=10^{11}-10^{13}M_{\odot}$ \citep{Ouchi03}. Furthermore, \citet{WL06} have recently presented a simple model which reproduces the observed Ly$\alpha$ luminosity function of \citet{Ka06}. In their model, the observed LAEs have $M_{\rm tot}=10^{10}-10^{11} M_{\odot}$. Using detailed semi-analytic galaxy formation models \cite{LD06} found the typical halo mass of the observed LAEs to be $M_{\rm tot} \sim 10^{11} M_{\odot}$. For the important results presented in this paper, our analysis is repeated for the case $M_{\rm tot}=10^{10}M_{\odot}$. For larger $M_{\rm tot}$, the infall region is more extended and the IGM transmission is expected to be (even) lower. We assumed a star formation rate of $\dot{M}_*=10M_{\odot}$/yr, and the escape fraction of ionising photons to be $f_{\rm esc}=0.1$. The spectrum of the galaxy is of the form $J(\nu) \propto \nu^{\beta}$, where $\beta=-4.5$, for $\nu \geq \nu_H$ (\S~\ref{sec:ew}). The impact of clustered sources on the ionising background is discussed later [here, $B_{\rm c}(r)=1$, Eq~\ref{eq:gamma}]. According to Eq~(\ref{eq:llya}-\ref{eq:qdot}), this galaxy has an intrinsic Ly$\alpha$ luminosity of \llya$=2\times 10^{43}$ erg s$^{-1}$.

In Figure~\ref{fig:fig1} we show the shape of the Ly$\alpha$ line after it has been processed by the IGM. The {\it solid}/{\it dashed lines} show the observed/intrinsic Ly$\alpha$ spectrum, respectively. As discussed in \S~\ref{sec:intro}, the blue side of the line ($x>0$) is suppressed significantly by the IGM. However, the suppression extends into the red part of the line. This is because of resonant absorption by residual neutral hydrogen gas that is falling onto the galaxy.  The peak of the observed spectrum is redshifted relative to the true line center by an amount set by the infall velocity. Here, the redshift is $v_{\rm IGM}(r_{\rm vir})=-v_{\rm circ}$.

We denote the flux density in the Ly$\alpha$ line by $J(x)$, where $x$ is the dimensionless frequency variable introduced in \S~\ref{sec:igmtrans}. The total transmission is then given by
\begin{equation}
\mathcal{T}_{\alpha}\equiv\frac{\int_{-\infty}^{\infty}dx J(x)\langle e^{-\tau}\rangle(x)}{\int_{-\infty}^{\infty}dx J(x)},
\end{equation} where $\langle e^{-\tau}\rangle(x)$ is given by Eq~(\ref{eq:trans}). For the example shown in Figure~\ref{fig:fig1}, $\mathcal{T}_{\alpha}=0.12$, i.e. only $12\%$ of the flux that emerges from the galaxy is transmitted through the IGM. The total observed flux of this galaxy is then $\mathcal{T}_{\alpha}\times $\llya$/[4\pi d^2_L(z)]=6\times10^{-18}$ erg s$^{-1}$ cm$^{-2}$. 

The exact shape of the observed Ly$\alpha$ line and the total transmitted flux depends on several other parameters, such as the assumed width of the intrinsic width of the Ly$\alpha$ line; the total ionising luminosity of the galaxy; the value of the ionising background etc. The variation in observable Ly$\alpha$ properties with these, and other parameters is studied next.
\subsection{Variation of the Line Shape}
\label{sec:param}
\begin{figure*}
\vbox{\centerline{\epsfig{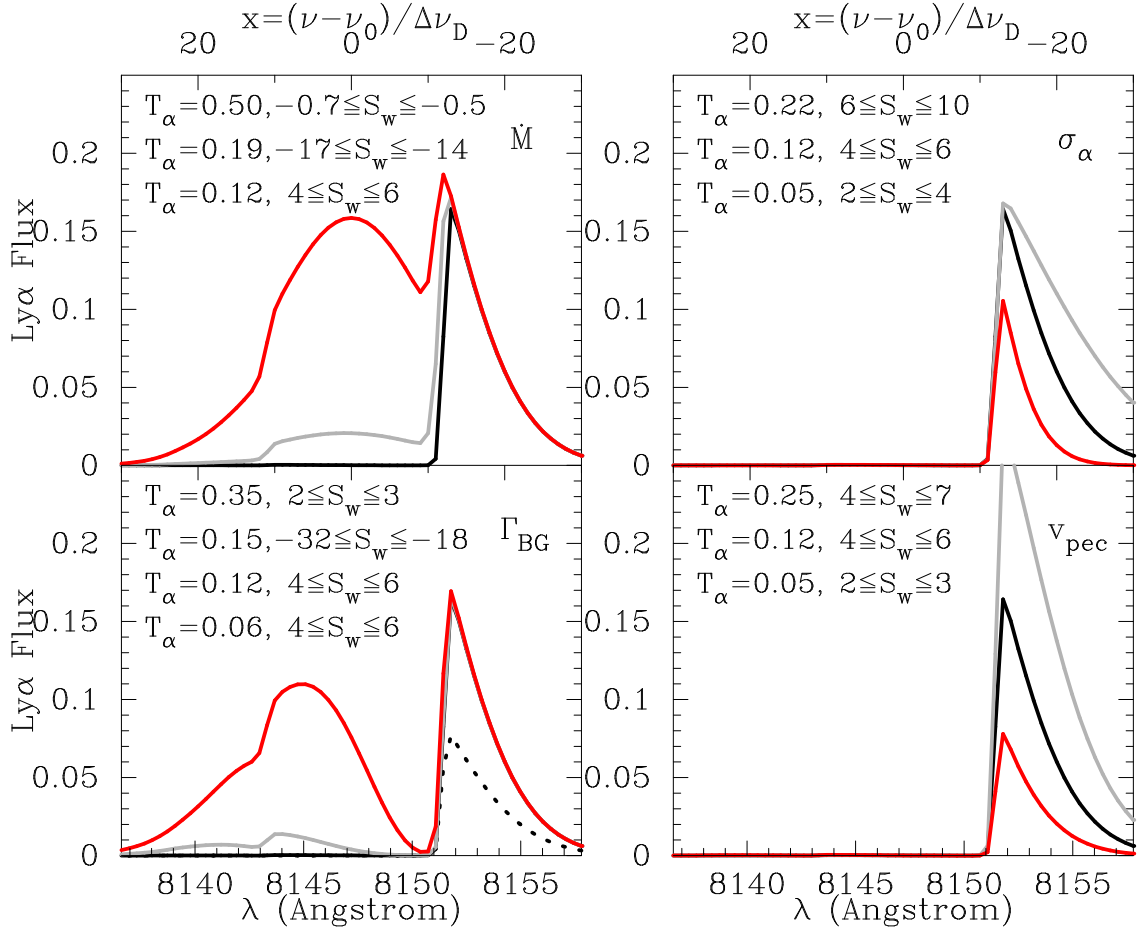}}}
\caption[]{The observed Ly$\alpha$ line from $z=5.7$ galaxies in a reionised IGM for a range of different models. The upper right corner of each panel shows the model-parameter that is varied. {\it Top Left:} $\dot{M}_*=10M_{\odot}$/yr ({\it black}), $\dot{M}_*=10^2M_{\odot}$/yr ({\it grey}) and  $\dot{M}_*=10^3M_{\odot}$/yr ({\it red}). {\it Top Right:} $\sigma_{\alpha}=1.0v_{\rm circ}$ ({\it black}), $\sigma_{\alpha}=0.7v_{\rm circ}$ ({\it red}) and $\sigma_{\alpha}=1.5v_{\rm circ}$ ({\it grey}). {\it Lower Left}: Variation of the ionising background: no ionising background ({\it dotted}), $\Gamma_{\rm BG}$({\it black}), $10\Gamma_{\rm BG}$ ({\it grey}) and $100\Gamma_{\rm BG}$ ({\it red}). {\it Lower Right}: Impact of galaxies peculiar velocity: $v_{\rm pec}=0$ ({\it black}), $v_{\rm pec}=-0.5\sigma_{\alpha}$ ({\it red}) and $v_{\rm pec}=0.5\sigma_{\alpha}$ ({\it grey}). For each model the total transmission, $T_{\alpha}$ and the skewness of the line, $S_W$, are shown.}
\label{fig:fig2}
\end{figure*}

In Figure~\ref{fig:fig2} we show the impact of various parameters on
the shape of the Ly$\alpha$ line after processing through the IGM.
 The IGM transmission for each curve is shown as a label. Also shown is the 'skewness parameter', $S_W$ \citep{Ka06}\footnote{Here, we denote the flux density in pixel number 'i' as $f(x_i)$. The parameter $S_W$, which has units of \AA, is defined as $S_W\equiv [\lambda_{10,{\rm red}}-\lambda_{10,{\rm blue}}]S$, where $\lambda_{10,{\rm red/blue}}$ are the observed wavelengths at which the flux drops to 10\% of its peak value, and $S=\frac{1}{I\sigma^3}\sum_{i=1}^n(x_i-\bar{x})^3f(x_i)$, where $I=\sum f(x_i)$, $\bar{x}=\sum x_i f(x_i)/I$ and $\sigma=[\sum (x_i-\bar{x})^2 f(x_i)/I]^{1/2}$. For a detailed definition and discussion of the skewness parameter, the reader is referred to the Appendix of \citet{Ka06}.}. This parameter quantifies the asymmetry of an emission line, which can be used to separate high redshift LAEs from lower redshift interlopers. This separation is possible as absorption in the IGM causes high redshift Ly$\alpha$ emission lines typically to be asymmetric (with a positive/negative skewness when the spectrum is shown as a function of wavelength/frequency, respectively), while lower redshift H$\alpha$, [OIII] or [OII] emitters are not. \citet{Ka06,Shima06} found that the requirement $S_W \geq 3$ \AA\hs can effectively distinguish between the two populations. Calculating $S_W$ for our models is not trivial. The largest contribution to the skewness comes from pixels that lie far from the observed Ly$\alpha$ peak flux, and we found that the precise value of $S_W$ depends somewhat on the assumed flux threshold, $f_{\rm min}$, below which the Ly$\alpha$ line is lost in the noise. In particular, a very low level of residual flux on the blue side of the Ly$\alpha$ line can cause $S_W$ to be negative for some of our models. To account for the dependence of $S_W$ on $f_{\rm min}$, we have shown the values of $S_W$ we found for each model for the range $0.01f_{\rm max} \leq f_{\rm min} \leq 0.1f_{\rm max}$, where $f_{\rm max}$ is the observed peak Ly$\alpha$ flux density.
\begin{itemize}
\item {\it Top Left:} Variation of $\dot{M}_*$: Here, we have plotted the cases $\dot{M}_*=10M_{\odot}$/yr ({\it black}), $\dot{M}_*=10^2M_{\odot}$/yr ({\it grey}) and  $\dot{M}_*=10^3M_{\odot}$/yr ({\it red}). Increasing $\dot{M}_*$ results in a higher output of ionising photons which reduces the neutral fraction in the infall region, and allows more flux, both red and blueward of the line, to be transmitted. Note that for $\dot{M}_* \geq 100 M_{\odot}$/yr) $S_W <3$ \AA\hs in difference to known high-redshift LAEs. For $\dot{M}_*=10^2M_{\odot}$/yr the spectrum shows a long tail on the blue side of the line, which makes $S_W$ negative. On the other hand, for $\dot{M}_*=10^3M_{\odot}$/yr the spectrum becomes practically symmetric with a skewness parameter $-0.7 \lsim S_W \lsim 0.5$ \AA. The precise value of $\dot{M}_*$ at which $S_W$ becomes less than 3 \AA\hs depends on the size of the infall region, and therefore on $M_{\rm tot}$. For example, given $M_{\rm tot}=10^{10}M_{\odot}$, $S_W \lsim 3$ \AA\hs for $\dot{M}_*\gsim 10 M_{\odot}$/yr. 
\item {\it Top Right:} Variation of the width of the intrinsic Ly$\alpha$ line: $\sigma_{\alpha}=1.0v_{\rm circ}$ ({\it black}), $\sigma_{\alpha}=0.7v_{\rm circ}$ ({\it red}) and $\sigma_{\alpha}=1.5v_{\rm circ}$ ({\it grey}). Clearly, increasing $\sigma_{\alpha}$ increases $\mathcal{T}_{\alpha}$, because a larger fraction of photons lie redward of the wavelength range affected by the infalling gas. Furthermore, the sharp break in the profile becomes more prominent towards larger $\sigma_{\alpha}$.
\item {\it Lower Left:} Variation of the ionising background $\Gamma_{\rm BG}$: We show cases in which the ionising background is $\Gamma_{\rm BG}$ ({\it black}), $10\Gamma_{\rm BG}$ ({\it grey}) and $100\Gamma_{\rm BG}$ ({\it red}). The main effect of raising the intensity of the ionising background is to increase the transmission on the blue side of the line. The skewness parameter, $S_W$, varies non-monotonically for similar reasons as those mentioned above: because of the long blue tail in the spectrum, $S_W$ is very negative for the case of $10\Gamma_{\rm BG}$, while the spectrum becomes almost symmetric again for the case of $100\Gamma_{\rm BG}$. The total IGM transmission depends only weakly on the intensity of the ionising background. The reason is that the transmission is determined strongly by the neutral fraction in the infalling gas, which is set primarily by the strength of the ionising luminosity of the galaxy. This can be understood as follows: define the `proximity radius', $r_{\rm prox}$, as the radius at which the photoionisation rate due to the galaxies ionising photons equals that of the ionising background ($\Gamma_{\rm BG}=\Gamma_{\rm Gal}$). For the fiducial model, the proximity radius lies at $r_{\rm prox} \sim 200 $ kpc ($\sim 10 r_{\rm vir}$), which is well outside the infall region. For $r \ll r_{\rm prox}$, the ionisation rate due to ionising background makes up only $\sim (r/r_{\rm prox})^2 \sim [0.01-0.25]$ of the total ionisation rate within the infall region [in which $r \sim (1\rightarrow 5)r_{\rm vir}$, see \S~\ref{sec:rho}]. As a result, raising $\Gamma_{\rm BG}$ by a factor of $\sim 10$ only contributes significantly to $\Gamma_{\rm tot}$ in the outer layers of the infall region.

Also shown is the spectrum that would be observed  when the ionising background is absent, $\Gamma_{\rm BG}=0.0$ ({\it dotted line}). The damping wing of the neutral IGM (outside of the HII sphere created by this galaxy), significantly reduces the flux redward of the infall feature as well. This damping wing optical depth is only important when the neutral fraction in the IGM is large $x_{\rm HI} \gsim 0.1$ \citep{Santos04}\footnote{The size of the Stromgren sphere was calculated from $\dot{Q}_H \times t_{\rm gal}=\int_{r_{\rm vir}}^{R_s}4\pi r^2 n_H(r)dr$, in which $t_{\rm gal}=3\times 10^8$ yr, is the life time of the galaxy. The neutral fraction in the IGM for $r >R_s$ as set to $1$, while for $r< R_s$ photoionisation equilibrium was assumed (Eq~\ref{eq:xh}).}
\item {\it Lower Right:} Impact of galaxy peculiar velocity: $v_{\rm pec}=0$ km s$^{-1}$({\it black}), $v_{\rm pec}=-0.5\sigma_{\alpha}$ ({\it red}) and $v_{\rm pec}=0.5\sigma_{\alpha}$ ({\it grey}). A negative peculiar velocity corresponds to the galaxy receding from the IGM. Clearly, more flux is transmitted towards larger negative velocities. The choice $v_{\rm pec}=-v_{\rm circ}$ is equivalent to models that use a static IGM, and would result in $\mathcal{T}_{\alpha}\sim0.5$. Furthermore, the shape of the Ly$\alpha$ line is quite insensitive to $v_{\rm pec}$: all lines are asymmetric with a sharp cut-off on the blue side of the line.
\end{itemize}
Lastly, we investigate is the impact of the radial dependent enhancement of the ionising background due to clustered sources. We showed in \S~\ref{sec:clustering} that this enhancement, denoted by $B_{\rm c}(r)$, can be as large as $10^2$, particularly in the close vicinity of the galaxy. In Figure~\ref{fig:impactcl} we show the observed Ly$\alpha$ line for the profile $B_{\rm c}(r)$ presented in the top panel of Fig.~\ref{fig:prox} ({\it black solid line}, $\lambda=1$ Mpc). The {\it red dashed line} represents the fiducial case with $B_{\rm c}(r)=1$ for comparison.
\begin{figure}
\vbox{\centerline{\epsfig{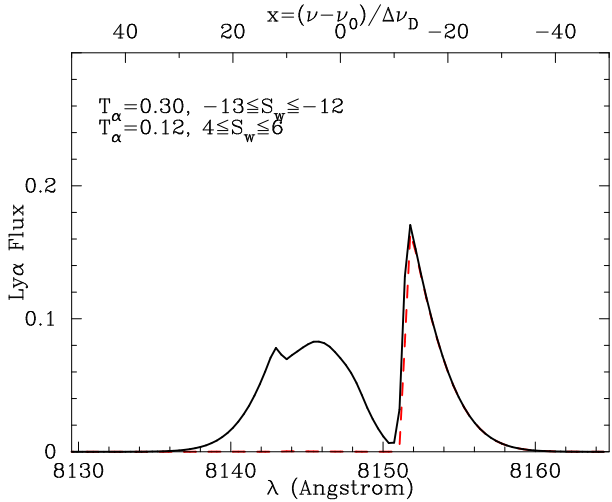}}}
\caption[]{The impact of an enhanced ionising background due to clustered sources surrounding a galaxy on its observed Ly$\alpha$ line. The {\it red dashed line} is the fiducial case with $B_{\rm c}(r)=1$, whereas the {\it black solid line} shows the case in which $B_{\rm c}(r)$ takes the form shown in Fig.~\ref{fig:prox} (the case of $\lambda=1$ Mpc). The blue side of the line is significantly enhanced, resulting in an increased transmission, $\mathcal{T}_{\alpha}=0.30$. Note that because of the skewness of $-13$\AA\hs$\leq S_w \leq-12$\AA, this LAE would have been misidentified a low redshift interloper (if the Ly$\alpha$ line were its only observed feature, see \S~\ref{sec:sym}).}
\label{fig:impactcl}
\end{figure} The figure shows that the blue side of the line is significantly enhanced (as we noticed above, see the lower left panel of Fig~\ref{fig:fig2}), resulting in a more prominent 'dip' in the spectrum. The overall transmission is enhanced by almost a factor of $3$ to $\mathcal{T}_{\alpha}=0.30$. The skewness of the line is $-13$\AA\hs$\leq S_w \leq-12$\AA, and therefore, this line would not be identified as a high-redshift LAE (see \S~\ref{sec:sym} for further discussion). It is interesting to note that since $\lambda$ decreases with redshift, the boost in the ionising background increases with redshift, which in turn implies that more Ly$\alpha$ lines may have $S_W \lsim 3$\AA\hs towards higher redshift.

We have found the observed shape of the Ly$\alpha$ line to vary widely, depending on $\dot{M}_*$, $\Gamma_{\rm BG}$, $\sigma_{\alpha}$ and the systemic velocity of the line, $v_{\rm pec}$. The main results can be summarised as follows: 

({\it i})  For large star formation rates and strong boosts of the ionising background by clustering of nearby sources, the Ly$\alpha$ line loses its trademark asymmetry (the skewness parameter for these models is $S_W \lsim 3$ \AA). This may become more common towards higher redshift. \citet{Taniguchi05,Ka06} found 20-30\% of their LAE-candidates to have symmetric lines. It could be that at least some of these are LAEs, as opposed to [O II] $\lambda 3727$ emitters. This is discussed in more detail in \S~\ref{sec:sym}.

({\it ii}) For most models, however, the Ly$\alpha$ line is asymmetric with a sharp cut-off on the blue side, and skewness parameters
 of $S_W \gsim 3$ \AA. In such cases, the IGM scatters more than half of the Ly$\alpha$ flux out of our line of sight, with mean transmissions of typically $\mathcal{T}_{\alpha}=0.1-0.3$. Furthermore, infall causes the peak of the Ly$\alpha$ line to be redshifted relative to the true line center.

({\it iii}) The Full Width at Half Maximum (FWHM) of the line is $\lsim v_{\rm circ}$ for most models, apart from those with large $\dot{M}_*$, $v_{\rm pec}$, $\sigma_{\alpha}$ or $B_{\rm c}(r)$. Models with large $\dot{M}_*$ and/or $B_{\rm c}(r)$ have $S_W \lsim 3$ \AA, and so produce either broad symmetric lines or lines that are asymmetric with skewness parameters that are negative.
\subsection{The Mean IGM Ly$\alpha$ Transmission Around Galaxies}
\label{sec:trans}
The previous discussion focused on the change in the Ly$\alpha$ line shape and Ly$\alpha$ transmission as a function of star-formation rate, ionising background, the width of the intrinsic Ly$\alpha$ and the peculiar velocity $v_{\rm pec}$. These calculations focused on $z=5.7$ galaxies. Here, we investigate
the total transmitted flux for the fiducial model in an ionised IGM (i.e. $f_{\rm esc}=0.1$, $v_{\rm pec}=0$, $\sigma_{\alpha}=v_{\rm circ}$ and $Z=0.05$) at additional redshifts of $z=4.5$ and $z=6.5$ as a function of $\dot{M}_{*}$. Thus we have assumed that reionisation was completed by $z=6.5$, although evidence exists from high redshift quasars may suggest otherwise \citep{WL04b,MH04,MH06}. However, the analysis presented here also applies when galaxies at $z=6.5$ are embedded in large HII bubbles which are expected to exist prior to the end of reionisation (see \S~\ref{sec:z7} for a more detailed discussion). The impact of varying the other model parameters can be inferred from Figure~\ref{fig:fig2}.
\begin{figure*}
\vbox{\centerline{\epsfig{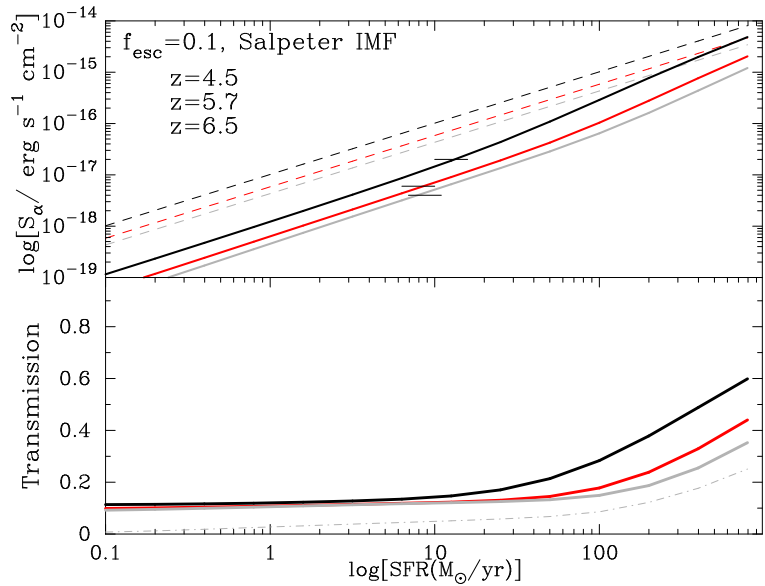}}}
\caption[]{{\it Bottom Panel}: Mean IGM transmission for a galaxy in a halo of mass $M_{\rm tot}=10^{11}M_{\odot}$ as a function of the star formation rate ($\dot{M}_*$), at $z=4.5$ ({\it black}), $z=5.7$ ({\it red}) and $z=6.5$({\it grey}). The {\it dotted grey line} represents the scenario in which the IGM is completely neutral at $z=6.5$. {\it Top Panel}: The observed Ly$\alpha$ flux for our fiducial model as a function of $\dot{M}_*$, at the same redshifts presented in the {\it bottom panel}. The {\it solid lines} and {\it dashed lines} show the observed flux when the IGM transmission is/is not accounted for respectively. For all SFRs the mean transmission is significantly less than 1, which may lead to an underestimate of $\dot{M}_*$. The thin {\it horizontal lines} denote the detection thresholds of recently performed Ly$\alpha$ surveys at $z=4.5$, $z=5.7$ and $z=6.5$ (see text).}
\label{fig:fig3}
\end{figure*}

The results of this calculation are shown in the lower panel of Figure~\ref{fig:fig3}. The {\it black}, {\it red} and {\it grey} lines represent galaxies at $z=4.5$, $z=5.7$ and $z=6.5$, respectively. The {\it grey dotted line} represents the scenario in which the ionising background is set to $0$ at $z=6.5$. The damping wing of hydrogen in the neutral IGM strongly suppresses the transmission as (we noticed already in \S~\ref{sec:param}). At all redshifts, $\mathcal{T}_{\alpha}\sim 0.1$ for $\dot{M}_* \lsim 10 M_{\odot}$/yr. For larger star formation rates, the mean transmission increases slowly and reaches the value $\mathcal{T}_{\alpha}=0.3$ only when $\dot{M}_* \sim 90 M_{\odot}$/yr, $\dot{M}_* \sim 300 M_{\odot}$/yr and $\dot{M}_* \sim 500 M_{\odot}$/yr at $z=4.5$, $z=5.7$ and $z=6.5$, respectively. Figure~\ref{fig:fig3} strongly suggests that not properly accounting for scattering in the IGM, can lead to significant underestimates of \llya, and thus of the star formation rates. In the upper panel, we show the observable Ly$\alpha$ flux as a function of star formation rate. The {\it solid/dashed lines} represent calculations in which the transmission was/was-not accounted for. The thin {\it horizontal lines} denote the detection thresholds of recently performed Ly$\alpha$ surveys at $z=4.5$ \citep{Dawson04}, $z=5.7$ \citep{Shima06} and $z=6.5$ \citep{Taniguchi05}. As an example, for a given Ly$\alpha$ flux from a $z=5.7$ galaxy, say $S_{\alpha}=10^{-17}$ erg s$^{-1}$ cm$^{-2}$, ignoring the IGM would lead to an estimated star formation rate of $\dot{M}_*=2M_{\odot}$/yr, while the actual star formation rate was $\dot{M}_* \sim 20M_{\odot}$/yr. Therefore, not accounting for IGM scattering could lead to substantial underestimates of the Ly$\alpha$-derived star formation rates occurring in LAEs. We discuss this in more detail in \S~\ref{sec:sfr}.
\subsection{Ly$\alpha$ Transmission in the Absence of a Virial Shock}
Our model assumes that no absorption takes place inside the virial radius.
The motivation for this assumption is that the virial temperature is well above $10^4$ K for the halos considered in this paper, and as a result, the virialised gas is collisionally ionised to a very high level (in our models the collisional ionisation rate of hydrogen in the virialised gas is $\sim 10^{-10}$ s$^{-1}$). However, the cooling time of this gas is short. In particular, the cooling time is shorter than the dynamical time of the system \citep{H00}, and hence the gas rapidly loses its pressure support and collapses further. Indeed, it has been argued that because of this efficient cooling, the virial shock does not exist in halos with masses below $M \lsim 2\times 10^{11}M_{\odot}$ (the precise mass depends on the metallicity of the gas, Birnboim \& Dekel, 2003). We may therefore expect that infall continues at radii less than the virial radius. The precise infall velocity profile of this gas is not well known. However, the absence of a virial shock would reduce the Ly$\alpha$ transmission further, especially in cases where the infall within the virial radius occurs at speeds exceeding $v_{\rm circ}$, because then, the absorption would extend even further into the red part of the line.

\section{Discussion}
\label{sec:discuss}

\subsection{Deriving Star formation Rates from Ly$\alpha$ Fluxes}
\label{sec:sfr}

In \S~\ref{sec:trans} we have shown how Ly$\alpha$ scattering in the IGM can reduce the detected Ly$\alpha$ flux by up to an order of magnitude. This implies that estimates of the star formation rate based on Ly$\alpha$ fluxes generally underestimate the true star formation rate. \citet{Shima06} found 28 spectroscopically confirmed LAEs. They found that the Ly$\alpha$ derived star-formation density for their survey volume lies a factor of $\sim 3$ below the estimates of \citet[][down to the same UV-luminosity]{Bouwens06}. Recently, \citet{Taniguchi05} found that the Ly$\alpha$ derived star formation rates lie, on average, a factor of $\sim 5$ below values derived from the UV-continuum measurements. The typical UV-derived star formation rates for galaxies in their sample were in the range $\sim 5-45 M_{\odot}$/yr, which according to our Figure~\ref{fig:fig3} would translate to $\mathcal{T}_{\alpha}\sim 0.1-0.2$. Scattering in the IGM therefore fully accounts for the apparent discrepancy between UV and Ly$\alpha$ derived star formation rates. 

Note however, that the discrepancy between UV and Ly$\alpha$ derived star formation rates, is not considered a serious issue. The reason for this is that Ly$\alpha$ photons may be subject to a different amount of dust attenuation than UV-continuum photons. For example, scattering of Ly$\alpha$ photons through the ISM increases their total traversed path through the galaxy, making Ly$\alpha$ photons more vulnerable to dust attenuation (on the other hand, in the case of a multiphase ISM, the Ly$\alpha$ may be subject to less attenuation, see \S~\ref{sec:ew}). Therefore, the observed discrepancy between Ly$\alpha$ and UV-derived star formation rates can also be interpreted as resulting from Ly$\alpha$ being preferentially destroyed by dust inside the galaxy \citep[see][for a discussion on the difficulties associated with estimating star formation rates from Ly$\alpha$ fluxes]{Kunth98}. However, it is interesting that our low value of the transmission through an ionised IGM provides a natural explanation for the discrepancy.

\subsection{Constraints from the Observed Ly$\alpha$ Line Width+EW.}
\label{sec:wind}

Figures~\ref{fig:fig1}-\ref{fig:impactcl} show that the typical observed line width of our model LAEs is $\sim 2-10$ \AA, with the fiducial model having a FWHM of $\sim 3$ \AA. The bulk of the observed LAEs in \citet{Shima06} have a FWHM in the range $\sim 2-14$ \AA, with the majority of the LAEs having FHWM$\gsim 5 \AA$. These values of the FWHM can be explained with larger $v_{\rm pec}$ or $\sigma_{\alpha}$. A larger star formation rate, or a boost in the ionising background have the same effect, but probably not while maintaining $S_w \gsim 3$ \AA. In particular, the high end of the range of observed FWHM cannot be explained this way. An elegant explanation for the large line width systems is provided by back-scattering off a superwind driven shell of neutral gas \citep{Ahn03,Ahn04}. Here, Ly$\alpha$ emitted by the galaxy scatters off an thick neutral shell that is receding from the galaxy with a velocity $v_{\rm exp}$. Photons that are scattered back to the galaxy have now picked up a redshift of $2v_{\rm exp}$, and may make it through the shell without being scattered. \citet{Ahn03} have shown that the Ly$\alpha$ line can easily be broadened by several times $v_{\rm exp}$. This back-scattered Ly$\alpha$ would all be on the red-side of the line center, and would thus be transmitted through the IGM. 

\citet{Ka06} have found that marginal evidence exists for an anti-correlation between the Ly$\alpha$ FWHM and the derived Ly$\alpha$ luminosity, i.e. the measured Ly$\alpha$ flux. This is contrary to what one would expect for the superwind scenario sketched above (in which the IGM transmission is larger) unless the wind-driven shells contain dust which could reduce the escape fraction of Ly$\alpha$ photons from the galaxy. An alternative explanation for this observation may be that that the broadest lines are broadened by resonant scattering (although we find this more unlikely as discussed in \S~\ref{sec:lyashape}).

\subsection{Comparing the Ly$\alpha$ and LGB Luminosity Functions.}

\citet{Ka06} compared the restframe-UV luminosity function of their $z=6.5$ LAE sample with that of the i-drop-out sample of \citet{Bouwens06} and found that the faint end slope for LAEs at $z=5.7$ \citep{Hu04,Shima06} and $z=6.5$ \citep{Ka06} is flatter than that of the i-drop out galaxies. The differences between the LAE and i-drop out UV-luminosity functions only becomes prominent for $M_{UV}\gsim -20.5$. \citet{Ka06} suggest that this is most likely due to a combination of detection incompleteness at faint Ly$\alpha$ and UV-fluxes ($M_{UV}\gsim -20.3$ corresponds to the $3-\sigma$ detection threshold). Using the standard conversion from UV-continuum flux density to star formation rate \citep[e.g.][]{K98}, we find that $M_{UV}\gsim -20.5$ corresponds to $\dot{M}_*\lsim$ $8.0 M_{\odot}$ yr$^{-1}$. According to Figure~\ref{fig:fig3}, the IGM transmission associated with these star formation rates is low, $\mathcal{T}_{\alpha} \lsim 0.12$ in this case, and the expected Ly$\alpha$ flux is $S_{\alpha}\lsim 6 \times 10^{-18}$ erg s$^{-1}$ cm$^{-2}$; almost exactly the detection threshold. For lower star formation rates (and higher $M_{\rm UV}$) a detection in Ly$\alpha$ requires a larger IGM transmission around these galaxies, which requires a large $v_{\rm pec}$, or $\sigma_{\alpha}$ (\S~\ref{sec:param}). This suggests that at fainter Ly$\alpha$ (and UV) fluxes indeed only a small subset of LAEs is currently detected.

\subsection{Symmetric and Double Peaked Ly$\alpha$ Emission Lines; A Source of Selection Bias?}
\label{sec:sym}

In \S~\ref{sec:param} we found that the Ly$\alpha$ line becomes either symmetric or asymmetric in the wrong direction ($S_W\lsim 3$ \AA) for large star formation rates and for strong enhancements of the ionising background. This suggests that the current criterion of asymmetry introduces biases against detection of galaxies undergoing intense episodes of star formation and/or embedded in regions of enhanced ionising background. In practice however, the criterion $S_W> 3$\AA\hs is used in combination with the broad band photometry in the B and V-bands \citep[e.g.][]{Shima06}. High-redshift LAEs are practically invisible in both these bands due to absorption in the IGM blueward of the rest-frame Ly$\alpha$ frequency. Therefore, only those objects having both $S_W< 3$\AA\hs and a detection in the B or V Band are classified as low redshift interlopers. Objects for which $S_W< 3$\AA\hs that are not detected in B and V are classified as an "unclear" objects. Approximately $15\%$ of the objects detected by \citet{Shima06} and \cite{Ka06} were classified as such and could thus be high-redshift LAEs. Six out of 34 of the $S_w > 3$ \AA\hs objects from \citet{Shima06} are bright in B or V. \citet{Shima06} argue that these objects are genuine LAEs, with the B/V flux contributed by nearby faint, contaminating sources. Likewise, some of the 19 objects in their 63-object spectroscopic sample with symmetric lines and detection in B/V could be LAEs and foreground B/V sources. However, the above considerations suggest a small fraction of $S_W< 3$\AA\hs LAEs and argue against a serious selection bias.

The low number of high-redshift LAEs with $S_W\lsim 3$ \AA\hs could be attributed to a small number of LAEs being embedded in regions of strongly enhanced ionising backgrounds, or to a small number of galaxies undergoing intense episodes of star formation. It could also imply that the IGM surrounding LAEs is denser than we have assumed. To make observed Ly$\alpha$ lines have $S_w > 3$ \AA, one needs to enhance the absorption in the IGM. Conversely, if we ignore infall, or lower the IGM density, then lower star formation rates or smaller enhancements of the ionising background are required to produce $S_W\lsim 3$ \AA-lines. We point out that although high-redshift LAEs with symmetric lines appear to be rare, these objects may be cosmologically very interesting. For example, recent simulations by \citet{Li06} show that the star formation rates in progenitor galaxies of the luminous $z=6$ quasars may be as large as $\dot{M}_{*}=10^4M_{\odot}$ yr$^{-1}$ (at $z=6-14$). Absorption in the IGM would almost certainly not produce $S_W\gsim 3$ \AA\hs Ly$\alpha$ lines.

Some Ly$\alpha$ emission lines shown in Figures~\ref{fig:fig3} and ~\ref{fig:impactcl} with $S_W\lsim 3$ \AA\hs have two peaks that are separated by $\sim 7$ \AA. Coincidentally, this is almost exactly the separation expected for a resolved [OII] doublet ($\sim6.6$ \AA\hs, in the observed frame at $z=1.2$). In cases where neither B nor V is detected, such a Ly$\alpha$ emitter may be confused with a $z=1.2$ [O II] doublet. Whether or not a Ly$\alpha$ line consists of two peaks depends on the following considerations. The largest contribution to the Ly$\alpha$ opacity at a frequency $x$ comes from the resonant region (defined as the region where the photons are at resonance, Eq~\ref{eq:tauH}). Therefore, if the resonant region lies a distance $r$ from the central galaxies, the total opacity to photons at frequency $x$ can be approximated by $\tau(x)\sim n_H(r)x_H(r)v_{\rm th }/[dv/dr]\propto n^2_H(r)/J(r)$. Here, we used the relation $x_H(r) \propto n_H(r)/J(r)$, and assumed for simplicity that the variation of $dv/dr$ with radius can be ignored. In our model, photons just blueward of the observed (red) peak, resonate with gas near the virial radius that falls inward at the maximum speed. If $n_H(r)$ decreases faster than $[J(r)]^{1/2}$ with increasing $r$, then the opacity decreases towards bluer wavelengths and produces a 'dip' in the observed spectrum. In our models $n_H(r)\propto r^{-1}$, while  to produce a dip requires $J(r)$ to change more slowly than $r^{-2}$, which is indeed the case when the ionising background is enhanced (Fig~\ref{fig:prox}). The precise location of the dip, and thus the separation of the two peaks, depends on the exact radial dependence of the density and velocity profiles of the surrounding gas, in addition to the ionising radiation field. Note that double peaked, spatially extended, Ly$\alpha$ emission has been observed \citep[e.g.][]{Ojik97,Wilman05}, often around radio galaxies. Various other mechanisms including superwinds and resonant scattering can create double peaked Ly$\alpha$ emission lines \citep[see][for a more detailed discussion on this issue]{V06}.  

\subsection{Ly$\alpha$ Transmission at Higher Redshifts: $z\geq 7$}
\label{sec:z7}
Our analysis has focused on the redshifts bins $z=4.5, 5.7$ and $z=6.5$, corresponding to windows of reduced atmospheric OH emission. Other windows exist at $z\sim 7$, $z=8$ and $z=8.8$ \citep{Nilsson06}. Ly$\alpha$ emission from higher redshifts is expected to be detected soon \citep[see][for a detection of a $z=7.0$ galaxy]{Iye06}. Our analysis is expected to be applicable here as well, even if reionisation was not completed until $z=6-6.5$. The reason is that prior to overlap, large bubbles (with typical sizes of $\sim 1-10$ physical Mpc, hereafter pMpc) of ionised gas are expected to populate the IGM, each of which may harbor many Ly$\alpha$ emitting galaxies \citep{F04a,F04b,WL05}. 

The analysis presented by \citet{Santos04} in which a galaxy sits inside its own HII region embedded in a neutral IGM, does not apply to most galaxies in this scenario. In particular, absorption by the damping wing of neutral intergalactic hydrogen is less important for galaxies inside these large HII regions. This is illustrated in Figure~\ref{fig:z8}. Here, we show the Ly$\alpha$ emission line of a galaxy at $z=8$ (the other model parameters, except for $\Gamma_{\rm BG}$, correspond to the fiducial model which is described in \S~\ref{sec:result}) for the case in which the galaxy is surrounded by its own HII region (which is $\sim 0.7$ pMpc for our model), and for cases in which the galaxy is embedded within a larger HII region created by an ensemble of nearby sources. The IGM outside the bubble is assumed to be neutral. The other cases shown are for $R_{\rm HII}=2.0$ pMpc ({\it red line}) and  $R_{\rm HII}=10.0$ pMpc ({\it grey line}).\footnote{We have assumed that the IGM is completely neutral outside the HII bubbles. Partial ionisation of the surrounding IGM \citep[e.g. by X-Rays,][]{Oh01,V01}, and neighboring bubbles reduce the contribution of the damping wing to the IGM transmission, in which case the dependence of the IGM transmission on HII bubble size is even smaller.} 

Figure~\ref{fig:z8} shows that damping wing absorption on the red side of the Ly$\alpha$ line center becomes decreasingly important towards larger $R_{\rm HII}$. However, resonant absorption the in infall region erases the Ly$\alpha$ line for $\lambda \lsim 1.095 \times 10^4$ \AA\hs in all cases. Since the transmission is regulated primarily by resonant absorption in the infall region and by damping wing absorption in the neutral IGM, $\mathcal{T}_{\alpha}$ is practically insensitive to the neutral fraction of hydrogen inside the bubble at $r \gsim 3-5 r_{\rm vir}$. The overall transmission for the three cases is $\mathcal{T}_{\alpha}=0.03$ ($R_{\rm HII}=0.7$ pMpc), $\mathcal{T}_{\alpha}=0.06$ ($R_{\rm HII}=2.0$ pMpc) and $\mathcal{T}_{\alpha}=0.10$ ($R_{\rm HII}=10.0$ pMpc). Therefore, at first glance, the presence of large HII bubbles prior to overlap boosts the  Ly$\alpha$ transmission by only a factor of $\sim 3$. We obtained similar results for a halo mass of $M_{\rm tot}=10^{10}M_{\odot}$, for cases in which the Ly$\alpha$ line is intrinsically narrower/broader ($\sigma_{\alpha}=0.7/1.5v_{\rm circ}$), and for cases in which the galaxy possesses a peculiar velocity with respect to the surrounding IGM ($v_{\rm pec}=\pm 0.5v_{\rm circ}$). However, we caution that when calculating the Stromgren radius, we (incorrectly) ignored recombinations in the HII region. Therefore, the Stromgren radius we calculate for our model is actually too large (although note that ionising radiation produced by a cluster of fainter, undetected, sources surrounding the galaxy may increase the Stromgren radius, e.g. Wyithe \& Loeb, 2005). If we decrease the Stromgren radius by a factor of $\sim 3$, then $\mathcal{T}_{\alpha}$ decreases by an additional factor of almost $2$. Furthermore, for galaxies in large bubbles the transmission could be boosted significantly due to nearby clustered sources (but note that this would likely result in Ly$\alpha$ lines with $S_W < 3$ \AA, see \S~\ref{sec:clustering}).

\begin{figure}
\vbox{\centerline{\epsfig{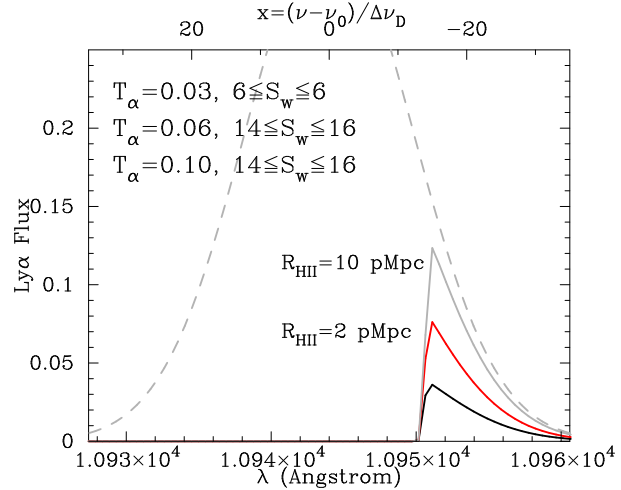}}}
\caption[]{The observed Ly$\alpha$ line of a $z=8.0$ galaxy (fiducial model parameters) for the case in which the galaxy carves its own HII region out of a fully neutral IGM ({\it black line}); when this galaxy lies in the center of a larger HII bubble of $2.0$ pMpc ({\it red line}); and $10.0$ pMpc ({\it grey line}). For each model $T_{\alpha}$ and $S_W$, are shown. The presence of large 'uber-HII region' boosts the Ly$\alpha$ transmission on the red side of the line. The {\it dashed grey line} represents the intrinsic spectrum.}
\label{fig:z8}
\end{figure}

The average HII-bubble size is related to the globally averaged ionised fraction of the universe by volume, $x_{\rm i,V}$ \citep{F06}, and therefore it is useful to calculate the IGM transmission as a function of HII-bubble size for various models. The result of this calculation is shown in Figure~\ref{fig:santos}, where we considered cases at $z=6.5$ ({\it dotted lines}) and $z=8.0$ ({\it solid lines}) for $\dot{M}_*=10M_{\odot}$ yr$^{-1}$ and $\dot{M}_*=10^2M_{\odot}$ yr$^{-1}$. The minimum radius for each line corresponds to cases of isolated galaxies in an otherwise neutral IGM. The relation between HII bubble size and neutral fraction depends on the nature of the ionising sources, the abundance of Lyman-limit systems, and on the redshift \citep{F06,M06}. An approximate relation between the characteristic HII bubble size $R_{\rm HII}$ and $x_{\rm i,V}$ derived by \citet{F06} is shown in the Figure for $0.4<x_{\rm i,v}<0.9$. The figure shows that the transmission drops only by a factor of $\sim 3$ compared to cases in which the IGM is fully reionised. As was mentioned above, part of the reason for this low contrast is that by ignoring recombinations within the Stromgren-sphere, we may have overestimated its size. Furthermore, clustered sources inside the large uber-HII bubbles could significantly boost the local ionising background, which would increase the transmission  (see \S~\ref{sec:clustering}).

\begin{figure}
\vbox{\centerline{\epsfig{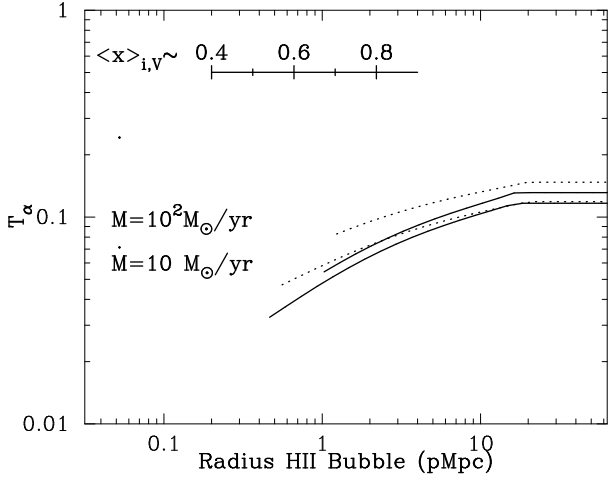}}}
\caption[]{The fraction of Ly$\alpha$ that is transmitted through the IGM as a function of HII-bubble size prior to the completion of reionisation. We have shown cases at $z=6.5$ ({\it dotted lines}) and $z=8.0$ ({\it solid lines}) for $\dot{M}_*=10M_{\odot}$/yr and $\dot{M}_*=10^2M_{\odot}$/yr. The minimum radius for each line corresponds to cases of isolated galaxies in an otherwise neutral IGM. The globally averaged ionised fraction of the universe by volume, $x_{\rm i,V}$, is related to the {\it characteristic} HII-bubble size \citep{F06}. An approximate relation between $R_{\rm HII}$ and $x_{\rm i,V}$ is also shown.}
\label{fig:santos}
\end{figure}

This result has the following implication: as reionisation of the IGM proceeds, HII bubbles grow larger in time until they overlap. The transmission of Ly$\alpha$ photons emitted by a galaxy inside such an HII bubble continuously grows with bubble size \citep{F06}. The precise dependence of $\mathcal{T}_{\alpha}$ on HII bubble size depends on the strength of the ionising background inside the bubbles. At overlap a `jump' in the ionising background may occur \citep[][although see Furlanetto \& Oh, 2005]{Gnedin00}. If such a jump occurs it will not result in a jump in the IGM transmission around LAEs, as we showed in \S~\ref{sec:param}. Therefore, LAEs of comparable luminosity should not appear to be significantly dimmer prior to overlap. The evolution in the Ly$\alpha$ luminosity function between $z=5.7$ and $z=6.5$ found by \citet{Ka06} is therefore not necessarily due to reionisation (unless the characteristic bubble size evolved very rapidly between $z=5.7$ and $z=6.5$, see Dijkstra et al. 2007 for a more detailed discussion). 

\section{Conclusions}
\label{sec:conclusions}

We have calculated the impact of an ionised intergalactic medium on the Ly$\alpha$ line emitted by galaxies at $z\geq 4.5$. Our model accounts for gas clumping in the IGM and for the fact that high-redshift galaxies typically reside in overdense regions, which causes the velocity field of the IGM to deviate from the Hubble flow. We calculated the observed shape of the Ly$\alpha$ line as a function of the galaxies ionising luminosity, the intrinsic width and systemic velocity of the Ly$\alpha$ line and the extra-galactic UV-background (which may be enhanced due to unseen clustered sources by a factor of $\sim 100$). 

We found that the observed shape of the Ly$\alpha$ line can display a wide range of shapes, depending on the precise values of $\dot{M}_*$, $\Gamma_{\rm BG}$, $\sigma_{\alpha}$ and  $v_{\rm pec}$ (see \S~\ref{sec:param} and Fig~\ref{fig:fig2}). Our most important findings are:

\begin{itemize}

\item  For large star formation rates, the strong asymmetry of the Ly$\alpha$ line (believed to be the defining characteristic of a high-redshift Ly$\alpha$ line) vanishes. For example, given $M_{\rm tot}=10^{11}M_{\odot}$ ($M_{\odot}=10^{10}M_{\odot}$), a star formation rate of $\dot{M}_*=100M_{\odot}$ yr$^{-1}$ ($\dot{M}_*=10M_{\odot}$ yr$^{-1}$) produces $S_w< 0$, whereas the value $S_W\geq 3$ \AA\hs has been shown to distinguish between confirmed high-redshift LAEs and lower redshift interlopers. Also, we find that $S_w\lsim 3$ \AA\hs for galaxies in which the ionising background was boosted significantly (which may become more common towards higher redshift) by surrounding clustered galaxies. An enhanced ionising background may occasionally result in a double-peaked Ly$\alpha$ emission line, mimicking a lower redshift [OII] emission line. The criterion $S_W\geq 3$ \AA\hs could therefore introduce a selection bias against galaxies with high star formation rates and/or galaxies embedded in an enhanced extragalactic UV-background. However, at z=5.7 and z=6.5 this is not a serious issue as only $\sim 15\%$ of the known Ly$\alpha$ emitter candidates has $S_W\lsim 3$ \AA. Nevertheless, recent cosmological simulations have shown that star formation rates as large as $\dot{M}_{*}=10^4M_{\odot}$ yr$^{-1}$ may occur at $z=6-14$ \citep{Li06}. Absorption in the IGM around these galaxies, while rare, would almost certainly not produce asymmetric Ly$\alpha$ lines.

\item In most models, the Ly$\alpha$ line is asymmetric ($S_w > 3$ \AA) with a sharp cut-off on the blue side. In these models, the IGM typically scatters more than half of the Ly$\alpha$ flux out of the line of sight, with a mean transmission of $\mathcal{T}_{\alpha}=0.1-0.3$.

\item The low values of the IGM transmission found in this paper suggest that not accounting for IGM scattering could lead to substantial underestimates of the Ly$\alpha$-derived star formation rates in LAEs. Although both the UV and Ly$\alpha$-derived star formation rates are subject to large uncertainties, our model naturally explains the apparent discrepancy that currently exists in the literature between the two.

\item Clustering of sources prior to overlap boosts the detectability of LAEs, as ensembles of clustered sources produce large HII bubbles. This reduces the impact the damping wing of the neutral IGM \citep{F06}. The precise dependence of $\mathcal{T}_{\alpha}$ on HII bubble size depends on the strength of the ionising background inside the bubbles. Furthermore, the possible boost in the ionising background that follows the overlap of HII bubbles (when reionisation is completed) barely increases the IGM transmission. Increasing the local ionising background by a factor of $\sim 10$ only increases the IGM transmission by a few per cent. This suggests that LAEs of comparable luminosity should not appear to be significantly dimmer prior to overlap.

\item The observed physical width of the Ly$\alpha$ line contains useful information. In particular, Our models have difficulties reproducing observed FWHM$\gsim 10$ \AA\hs, while conserving the asymmetry of the line, and we conclude that these LAEs are most likely to be galaxies with a superwind-driven outflow.

\end{itemize}

Our result that LAEs are not expected to become significantly brighter in a post-overlap IGM, naively implies that LAEs do not provide a direct probe of the end of reionisation. However, our work suggests that LAEs are not expected to dim significantly faster than $\propto d^{-2}_L(z)$ at $z \gsim 7$. We therefore expect Ly$\alpha$ surveys to be very fruitful in uncovering young galaxies at these higher redshifts, which will allow studies of star formation in the very first galaxies. Constraints on reionisation can then be obtained through various methods which include for example, cross-correlating the distribution of LAEs with intergalactic 21-cm emission \citep{WL06}; or by adding up the ionised volume surrounding individual LAEs as in \citet{MR06}. Moreover, with a large enough sample of LAEs at various redshift bins, it is possible to detect a weak evolution of the IGM transmission, which is expected to exist in our model due to the gradual growth of HII bubbles \citep{F06,M06,paperII}. The constraints on the IGM transmission can be further improved when Ly$\alpha$ luminosity functions are combined with the Ly$\alpha$ emitters' rest-frame UV luminosity functions \citep[as in][also see Dijkstra et al. 2007]{Ka06}. Additionally, since the abundance of observable emitters is modulated by the presence/absence of large HII bubbles, the two-point correlation function of LAEs may be used to constrain the dependence of $\mathcal{T}_{\alpha}$ on HII-bubble size \citep{F06,M06}, which puts constraints on the strength of the ionising background inside the bubbles.

Finally, we point out that our focus has been on the mean Ly$\alpha$ profile. In reality, fluctuations in the transmission from source to source are likely to exist. For example, the transmission is higher for sources viewed along a particularly low density path. It is highly unlikely that the majority of known LAEs are observed along such paths, as a high transmission is likely to be accompanied with $S_W \leq 3$ \AA. In future work we will address this scatter in the Ly$\alpha$ transmission in more detail using cosmological simulations.

{\bf Acknowledgments} JSBW and MD thank the Australian Research Counsel
for support and MD thanks the Center for Astrophysics for their hospitality. We thank Avi Loeb for useful discussions. We thank the anonymous referee for his/her careful reading of this manuscript and for a very helpful report that improved the content of this paper.

\appendix
\label{app:clustering}

\section{The Boost in the Ionising Background due to Clustering of Sources.}
We model the ionising background in the overdense regions surrounding massive galaxies using a semi-analytic model for biased galaxy formation near high redshift galaxies. Since we focus on $z>4$, beyond the peak redshift for quasar
activity, we consider the case where AGN do not contribute to the back-ground. Making the assumption that the ionising radiation generated by
galaxies was emitted over a Hubble time, and assuming the
ionising photon mean-free-path ($\lambda$) to be independent of density, 
we may derive an expression for the ionising background flux at radius $R$. 
We first calculate the ionising photon production rate per unit volume from
sources surrounding the central galaxies
\begin{equation}
\label{rate}
\frac{dN}{dt}(R)=\int_{M_{\rm min}}^{\infty}dM \left(\frac{4\pi D_{*}^3n_{\rm H}}{3H^{-1}}\right)\frac{dn\left(R,M_{\rm halo}\right)}{dM},
\end{equation}
where $n_{\rm H}$ is the hydrogen number density. In this expression the number density of halos surrounding the central halo $dn(R,M_{\rm
halo})/dM$ is found from the Press-Schechter mass function \citep{PS} with the modification of \citet{ST}, expressed in units of (physical Mpc)$^{-3}M_\odot^{-1}$. In the vicinity of a massive halo $dn/dM$ must be modified relative to the background universe. The dominant contribution to the modification of the halo mass function originates from the accelerated formation of halos in overdense regions. The calculation of this enhanced mass-function is described in \citet{WL05}. 

The quantity $D_*$ in equation~(\ref{rate}) is the characteristic physical size of a spherical HII region generated by stars in a galaxy with a dark matter halo of mass $M_{\rm halo}$, and was given by \citet{L05}
\begin{eqnarray}
\label{Dstar}
\nonumber
\nonumber D_{*} &=& 0.75\left(\frac{M_{\rm
halo}}{10^{10}M_\odot}\right)^{1/3}\left(\frac{f_{\star}f_{\rm esc}/N_{\rm
reion}}{0.003}\right)^{1/3}\\
&\times&\left(\frac{N_{\star}}{4000}\right)^{1/3}
\left(\frac{1+z}{7.5}\right)^{-1}\mbox{Mpc},
\end{eqnarray}
where $f_{\star}$ is the fraction of baryons in the halo that are turned
into stars, and $f_{\rm esc}$ is the fraction of ionising photons that
escape from the galaxy into the surrounding IGM.  The number of ionising
photons per baryon incorporated into Pop-II stars is $N_{\star}\sim 4000$ \citep{Bromm01}, and $N_{\rm reion}$ denotes the number of
photons required per baryon for the reionisation of the HII region. The
value of $N_{\rm reion}$ depends on the number of recombinations and hence
on the clumpiness of the IGM. 

When (equation~\ref{rate}) is substituted into Eq~(\ref{eq:boost}), the enhancement of ionising background at radius $R$ can be computed. Several model parameters need to specified: First of all, we take $M_{\rm min}$ to correspond to $v_{\rm circ}=50~{\rm km~s^{-1}}$, which reflects the fact that only
galaxies more massive than this can form in an ionised IGM, and hence contribute to the ionising background. Furthermore, we have assumed the ionising photons' mean-free-path to be equal to the measured value at $z=5.7$, $\lambda \sim1-5$ physical Mpc \citep{Fan06}. Last, we need to specify the precise dependence of $D_*$ on halo mass, by specifying the product $f_{\star}f_{\rm esc}/N_{\rm reion}$. These model parameters were chosen to achieve reionisation at $z=6$ as in \citet{WL05}.

\label{lastpage}
\end{document}